\documentclass[prb,twocolumn]{revtex4-1} 

\usepackage[breaklinks,colorlinks,citecolor=blue]{hyperref}

\usepackage{graphicx}

\usepackage[utf8]{inputenc}

\usepackage{amsfonts}
\usepackage{amssymb}

\newcommand{\be}{\begin{equation}}
\newcommand{\ee}{\end{equation}}
\newcommand{\bea}{\begin{eqnarray}}
\newcommand{\eea}{\end{eqnarray}}
\newcommand{\Dd}{\mathrm{d}}

\newcommand{\fvec}[1]{\underline{#1}}

\renewcommand{\doi}[1]{{\sc doi:} \href{http://dx.doi.org/#1}{\detokenize{#1}}}

\newcommand{\jName}[1]{#1}
\newcommand{\jVol}[1]{\textbf{#1}}

\usepackage{etoolbox}
\newtoggle{anonymizedVersion}
\settoggle{anonymizedVersion}{true}
\settoggle{anonymizedVersion}{false}

\newcommand{\makeAnon}[2]{\iftoggle{anonymizedVersion}{#1}{#2}}

\begin{document}

\title{Counterintuitive properties of relativistic relative motion for accelerated observers}

\author{\makeAnon{Name anonymized}{M P{\"o}ssel}}
\address{\makeAnon{Address anonymized}{Haus der Astronomie and Max Planck Institute for Astronomy, K{\"o}nigstuhl 17, 69124 Heidelberg, Germany}}
\email{\makeAnon{nemo@anonymized.org}{poessel@hda-hd.de}}

\begin{abstract}
A challenge in teaching about special relativity is that a number of the theory's effects are at odds with the intuition of classical physics, as well as student's everyday experience. The relativity of simultaneity, time dilation and length contraction are prominent examples. This article describes two additional, less well-known counter-intuitive properties, both of which follow from the relativistic definition of relative motion, in situations with accelerated observers: (a) when two objects have a constant radar distance and are by that standard ``at relative rest,'' their relative speed is not necessarily zero, and (b) for two observers A and B, a situation is possible where A considers the two to be approaching each other, while B considers them to be moving away from each other. In general relativity, the generalisations of these two properties prove helpful for understanding static gravitational fields, and they also provide some insight into the nature of horizons. This is a preprint version of the article M. P\"ossel, Am.\ J.\ Phys.\ 92, 957--964 (2024), \doi{10.1119/5.0203454}.
\end{abstract}
\maketitle 

\section{Introduction}
Relative velocities for inertial frames play an important role in special relativity, and thus feature prominently in all expositions of the theory. Velocities measured by non-inertial observers, on the other hand, are absent from most introductory accounts of special relativity. With the transition to general relativity, relative motion takes a back seat altogether: in that theory, relativity is not about relative motion, but more generally about covariance and the freedom to choose general coordinates. 

In this article, we argue that it is nonetheless instructive to explore the concept of non-inertial relative velocity in special relativity, as well as its counterpart in general relativity. In special relativity, such exploration leads to additional counter-intuitive properties of relativistic physics. In general relativity, it proves helpful for understanding static gravitational fields, notably the exterior Schwarzschild solution.

In the context of classical mechanics, most of us will have formed intuitive notions about relative motion that one might be tempted to regard as universal:
\begin{enumerate}
\item When the physical distance between two objects is constant, their relative speed is zero.
\item Objects in relative motion are moving either away from each other (positive radial velocity) or towards each other (negative radial velocity), but not both at the same time.
\end{enumerate}
Both of these statements are meant to refer to measurements --- this article is not concerned with apparent effects, e.g. how an idealised camera would record what is happening. 

As we shall see in sec.~\ref{Sec:RadarDistance} and sec.~\ref{Sec:Counterintuitive}, once we allow for acceleration, the straightforward special-relativistic generalisations of these statements both turn out to be wrong, providing two more examples for how relativity runs counter to classical intuition. In contrast with the standard examples for counter-intuitive relativistic effects, which are associated with inertial systems (relativity of simultaneity, time dilation, length contraction), the two examples presented here, concerning relative velocity, do not seem to be accounted for in the existing literature, although a similar effect was found by Rashidi and Ahmadi\cite{Rashidi2019} for how different observers measure acceleration. 

In sec.~\ref{Sec:Schwarzschild}, we examine how those properties can help us to better understand certain aspects of general relativity, in particular static spacetimes with gravitational redshift. The analysis also throws new light on standard descriptions of the equivalence principle and adds a new twist to understanding horizons.

\section{Relative velocity from four-velocities}
\label{Sec:SRRelativeVelocity}

We begin by reviewing the concepts of relative speed and relative radial velocity in special relativity. Four-vectors provide us with a coordinate-independent language that allows for statements about observers and objects in arbitrary motion. 

It is straightforward to obtain the velocity of one inertial frame ${\cal I}_1$ relative to another such frame ${\cal I}_2$: Describe the situation in the frame ${\cal I}_1$, and the ordinary three-space velocity assigned to the origin (or any other fixed point) of ${\cal I}_2$ in the frame ${\cal I}_1$ is the relative velocity in question. If we align the spatial axes of the two systems in the same way, the {\em reciprocity principle} holds: The velocity vector of ${\cal I}_2$ as expressed in ${\cal I}_1$ is the opposite of the velocity vector of ${\cal I}_1$ as expressed in ${\cal I}_2$. In consequence, the relative speed, defined as the magnitude of the relative velocity, is the same, regardless of which of the two systems we use as our reference frame. 

If instead of inertial frames, we instead want to describe the relative motion of two observers, or objects, 1 and 2, each of which is in arbitrary motion, more information is needed. After all, when the state of motion of either object changes, the relative velocity will, in general, change as well. Even in classical physics, in such a situation we would need to state at which moments in time we are evaluating a relative velocity. If, for instance, we want to calculate the Doppler effect for a specific signal, we would need to take into account the emitting object's state of motion at the time of emission, and the receiving object's state of motion at the time of reception of the signal in question, and determine the relative speed linking those two states. For a second signal that is moving from the first to the second object at a different speed, the result we obtain for the relative velocity that governs that signal's Doppler shift will in generally be different. In this sense, relative velocity is not a property that merely depends on which two objects we are talking about. Even in classical physics, there is an additional dependence on two moments in time, one for each object.

In special relativity, where simultaneity is relative, there isn't even a coordinate-independent way of talking about the relative velocity ``at a given moment in time'' anymore. In that setting, for any coordinate-independent description of relative motion, we will need to specify two events ${\cal E}_1$ and ${\cal E}_2,$ one on each object's world-line, in order to define relative speed. Only then do we have sufficient information to be able to talk about ``the speed of object 1 at  ${\cal E}_1$ relative to object 2 at ${\cal E}_2$.''

At this point,  there is a natural way of defining relative speed in special relativity: the speed $v$ of object 1 at  ${\cal E}_1$ relative to object 2 at ${\cal E}_2$ is defined as the speed of the inertial frame ${\cal I}_1$ that is momentarily co-moving with object 1 at  ${\cal E}_1$ relative to the inertial frame ${\cal I}_2$ that is momentarily co-moving with object 2 at  ${\cal E}_2$. Note that this cannot be generalised unambiguously to a fully-fledged relative velocity, as the three-vector components of such a velocity will still depend on the (arbitrary) spatial orientation of the two systems.

In the usual \mbox{(pseudo-)}Cartesian coordinates and with the Minkowski metric $\eta=\mbox{diag}(-1,+1,+1,+1),$ the four-velocity of object 1 at event ${\cal E}_1$ in the inertial frame ${\cal I}_2$ will have the form
\be
\fvec{u}=\gamma(v)\left(
\begin{array}{c}
c\\
v_x\\
v_y\\
v_z
\end{array}
\right),
\label{GeneralFourVelocity}
\ee 
with three-velocity components $v_x, v_y$ and $v_z$ and speed $v=\sqrt{v_x^2+v_y^2+v_z^2}$, where the gamma factor is
\be
\gamma(v)\equiv [1-(v/c)^2]^{-1/2},
\ee
and where we introduce the convention of denoting four-vectors by underlined letters. 
The three-vector $\vec{v} = (v_x,v_y,v_z)^T$ determines the velocity of the first object, evaluated at ${\cal E}_1$, relative to the second object, and corresponds to a (coordinate-system-dependent)  relative velocity. $v_{\mathrm{rel}}\equiv v$ is the relative speed, according to our definition.

Expression (\ref{GeneralFourVelocity}) suggests a well-known, coordinate-free short-cut for determining $v_{\mathrm{rel}}$ from the objects' four-velocities.  In its own momentary rest system ${\cal I}_2$, the four-velocity of object 2 is $\fvec{w}=(c,0,0,0)^T$. To find $v_{\mathrm{rel}}$, we need only calculate
\be
\gamma(v_{\mathrm{rel}}) = -\eta(\fvec{w},\fvec{u})/c^2
\label{RelativeSpeed}
\ee
and solve for $v_{\mathrm{rel}}$. This formula is coordinate-independent, and it is also symmetric in the two arguments, showing that the relative speed is reciprocal: Once the two objects and the two events are specified, we need not distinguish between the first object's speed relative to the second one, or vice versa. Our $v_{\mathrm{rel}}$ defines relative speed in a way that depends only on the objects 1, 2 and the events ${\cal E}_{1,2}$, just as it should be.

An interesting special case is purely radial motion. In terms of the momentarily co-moving inertial systems  ${\cal I}_{1,2}$, chosen here for simplicity so that ${\cal E}_{1,2}$ are in their system's spatial origin, this is equivalent to the two spatial origins moving directly away from or directly towards each other. We define the {\em relative radial velocity} $v_R$ so that when the origins are moving away from each other, $v_R\equiv +v_{\mathrm{rel}}$, whereas for motion towards each other, $v_R\equiv -v_{\mathrm{rel}}$.

From the description in terms of momentarily-comoving inertial system, it is clear that for purely radial motion, $v_R$ governs the longitudinal relativistic Doppler effect: with the usual abbreviation
\be
z=\frac{\lambda_r-\lambda_e}{\lambda_e},
\ee
where $\lambda_e$ is the wavelength of a light signal emitted by object 1 at the event ${\cal E}_1$, and $\lambda_r$ the wavelength of that same light signal as it is received by object 2 at the event ${\cal E}_2$, we have
\be
1+z = \sqrt{\frac{1+v_R/c}{1-v_R/c}},
\label{SRDoppler}
\ee 
so that a redshift $z>0$ corresponds to $v_R>0$ for the two objects, as evaluated at the emission and reception event, so the objects are moving away from each other, while a blueshift $z<0$ indicates negative radial relative velocity, corresponding to motion towards each other. Conversely, in situations where ${\cal E}_1$ and ${\cal E}_2$ are associated respectively with the emission and reception of a light signal, we can use the measured redshift to reconstruct $v_R$ using (\ref{SRDoppler}).

\section{Relative velocity for accelerated observers with constant radar distance}
\label{Sec:RadarDistance}

Next, let us consider accelerated observers and/or objects. There are numerous cases in which accelerated observers are used as helpful pedagogical tools, notably in the context of the so-called twin paradox\cite{Marsh1965,Dolby2001,Minguzzi2005a} and in connection with the equivalence principle, comparing the effects of constant acceleration and a homogeneous gravitational field.\cite{Desloge1989,Jones2008,Munoz2010}

\begin{figure}[htbp]
\begin{center}
\includegraphics[width=0.7\linewidth]{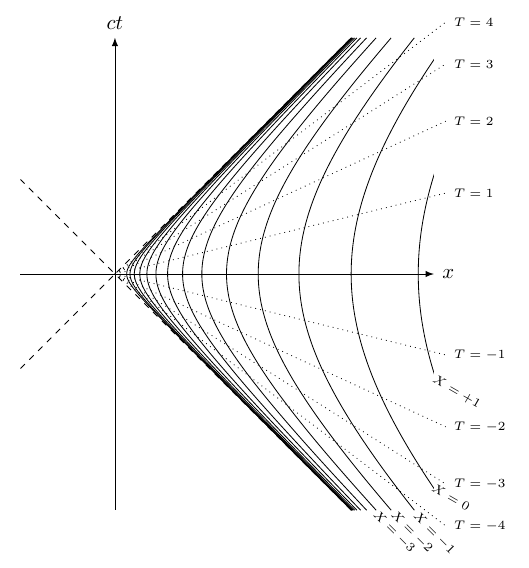}
\caption{Some worldlines for observers of the family specified by (\ref{LassCoordinates1}) and (\ref{LassCoordinates}) with constant acceleration parameter $a$, with equally-spaced values in $X$.  \label{Fig:LassSpacetime}}
\end{center}
\end{figure}

For simplicity, we consider only one spatial and one time dimension. In this $1+1$-dimensional spacetime, we examine a particular family of accelerated observers defined by Lass\cite{Lass1963} and Minguzzi:\cite{Minguzzi2005b} Let $a$ be a parameter with the physical dimensions of an acceleration, and $x,t$ coordinates in an inertial system ${\cal I}$. We then define a family of observers using two parameters $X,T$, where $X$ parametrises the family and $T$ each observer's world line, via 
\bea
\label{LassCoordinates1}
x &=& (c^2/a)\cdot e^{aX/c^2}\,\cosh(aT/c)\\[0.5em]
t &=& (c/a)\;\cdot   e^{aX/c^2}\,\sinh(aT/c).
\label{LassCoordinates}
\eea
Eliminating $T$ from equations (\ref{LassCoordinates1}) and (\ref{LassCoordinates}), we can see that in $\cal I$, each worldline is a hyperbola (cf.~Fig.~\ref{Fig:LassSpacetime}),
\be
x=c\sqrt{t^2+\left(\frac{c}{a}\right)^2\cdot\exp(2aX/c^2)}.
\label{XTrajectory}
\ee
Differentiation of this expression with respect to $t$ show that in ${\cal I}$, and thus in every other inertial reference frame, each of the observers is moving at sub-luminal speed, tending to $c$ only asymptotically as $t\to \pm\infty$.
If we include the whole range of $-\infty<X<\infty$, the worldlines fill out one of the two regions outside the light cone through the origin of $\cal I$. Within that region, the $X, T$ can serve as coordinates, and we can write the Minkowski line element $\Dd s^2$ in these new coordinates as
\be
\Dd s^2=\Dd x^2 - c^2\Dd t^2 = e^{2aX/c^2}\left[  dX^2-c^2\Dd T^2 \right].
\label{LassMetric}
\ee
Up to an overall ``conformal factor'' $e^{2aX/c^2}$, this is the Minkowski metric for $X, T$. Since $\Dd s^2=0$ is the condition that characterizes light-like world lines, in our $1+1$-dimensional spacetime, these world lines are given by
\be
X(T) = X_0 \pm c(T-T_0).
\label{LightPropagationXT}
\ee
This tells us explicitly that $X$ describes radar distances, as measured by our family of observers, and that in turn $T$ simultaneity follows the Einstein definition: $X,T$ are radar coordinates.\cite{Desloge1987} 

Since $\Dd s^2=-c^2\Dd\tau^2$, the proper time $\tau$ shown on the clock of one of the family of observers can be deduced from (\ref{LassMetric}) and the trajectory (\ref{XTrajectory}). Up to an arbitrary choice of origin, $\tau$ is related to the time coordinates $T$ and $t$ as 
\be
\tau = e^{aX/c^2}\,T =  e^{aX/c^2}\,\frac{c}{a}\,\sinh^{-1}\left(\frac{at}{c}e^{-aX/c^2}\right)
\label{ProperTimeRelation}
\ee
From this and the trajectory (\ref{XTrajectory}), it is straightforward to calculate the four-velocity of each such observer in ${\cal I},$ namely
\bea
\nonumber u^t \equiv \frac{\Dd (ct)}{\Dd\tau} &=& c \sqrt{
1+\left(
\frac{at}{c}e^{-aX/c^2}
\right)^2
}\\
&=&c\cdot\cosh(aT/c),\\[0.5em]
u^x\equiv\phantom{(c}\frac{\Dd x}{\Dd\tau} &=& at\cdot e^{-aX/c^2}
= c\sinh(aT/c).
\eea
This provides us with the input for calculating the relative speed (\ref{RelativeSpeed}) of arbitrary observers in our preferred family. With four-velocities $\fvec{u}_1,\fvec{u}_2$ for the two observers, and with the relative speed evaluated at time $T_1$ on the first observers's worldline and at the time $T_2$ on the second observer's, we obtain
\be
\gamma(v_{\mathrm{rel}}) = -\eta(\fvec{u}_1,\fvec{u}_2)/c^2 = \cosh(a[T_2-T_1]/c),
\ee
making use of the addition theorem for the hyperbolic cosine. Solving for the relative speed $v_{\mathrm{rel}}$, 
\be
v_{\mathrm{rel}} = c\cdot\tanh(a|T_2-T_1|/c).
\label{LassRelativeSpeed}
\ee

\section{Relative motion and accelerated observers}
\label{Sec:Counterintuitive}

Using the model situation described in the previous section, we can demonstrate the aforementioned new ways in which common intuition about relative speeds goes wrong for special-relativistic, accelerated objects.

\subsection{Relative speed vs.\ changes in physical distance}
\label{Sec:RelSpeedPhysDistance} 

If we were merely given the coordinates $X,T$ and the corresponding metric (\ref{LassMetric}), we might be tempted to argue as follows: Clearly, that metric is static, and the coordinates $X,T$ are adapted to that property --- none of the metric coefficients depends on $T$, and there are no mixed terms in $X$ and $T$. Thus, two objects that stay fixed at some given $X$ value, say, at $X_1$ and $X_2$, should be considered as being at rest relative to each other. 

\begin{figure}[htbp]
\begin{center}
\includegraphics[width=0.8\linewidth]{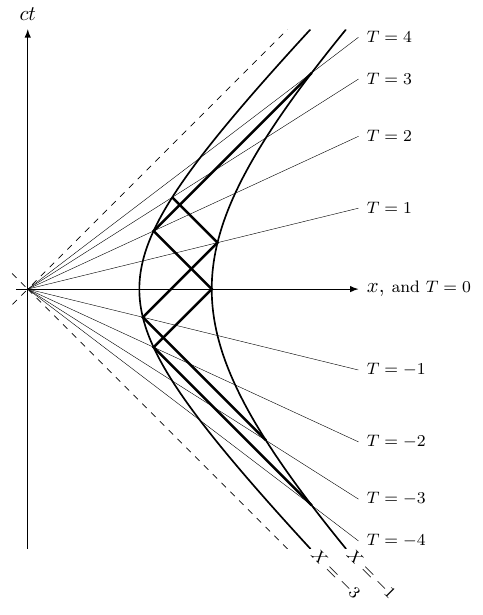}
\caption{Two of our accelerated observers ($X=-1$ and $X=-3$) exchanging light signals (thick diagonal $45^{\circ}$ worldlines). Also shown are nine lines of constant $T$}
\label{Fig:RadarDistances}
\end{center}
\end{figure}

If we distrust that line of argument because it relies on specific coordinates, we can go one step further: We let the observer at constant $X_1$ send light signals to the observer at constant $X_2$, receive reflected light signals back, and document their proper-time difference between the two. In our situation, an observer at $X_1$ would find that it always takes the same time for their light signals to travel to $X_2$ and back. From the external perspective of our original inertial system ${\cal I}$, this constancy is the result of a combination of different circumstances, which just happen to yield the same result every time. Fig.~\ref{Fig:RadarDistances} shows an example.
From the perspective of our inertial system, the worldline segment corresponding to light emission at $X=-3$ at time $T=-4$, and reception at $X=-1$ at $T=-2$, looks markedly different from, say, the worldline segment linking $X=-3$ at time $T=0$ and $X=-1$ at time $T=2$. In $X,T$ coordinates, on the other hand, light propagation really does follow the simple constant-speed relation (\ref{LightPropagationXT}). The corresponding radar distances determined by either observer are indeed constant.

Given that we are in special-relativistic spacetime, surely a constant radar distance like that is as close as we can come to making a physical statement about the distance between $X_1$ and $X_2$: the two observers {\em are} at rest relative to each other. Whenever the observer at $X_1$ repeats their radar measurement of the observer at $X_2$, they will get the same distance. The observer at $X_2$ could also perform the experiment, and would find a distance that does not change over time.

By these criteria --- constant static coordinate values, constant radar distance ---, the two observers at $X_1$ and $X_2$ are at rest relative to each other, at all times. But then there is the expression (\ref{LassRelativeSpeed}), which shows that while the two observers in question indeed have relative speed zero when evaluated at the same time, $T_1=T_2$, they have non-zero relative speed for comparisons at different times, and in particular at the pairs of events probed by sending a light signal in either direction. Each observer will see the other's signals to be Doppler-shifted. Being ``at (relative) rest,'' as determined by either criterion --- static coordinates, or constant radar distances -- is not a coordinate-independent property. Even for something that most readers will likely consider the best candidate for a ``physical distance'' we have in this situation, namely constant radar distance, constancy over time does not mean vanishing relative speed (\ref{LassRelativeSpeed}).

This unusual property is the counterpart to Bell's Spaceship Paradox:\cite{Dewan1959,Bell1987} In that case, two spaceships are accelerated identically as judged from an inertial system, which corresponds to relative speed zero at constant time in that inertial system. The result is a lengthening of physical distance as indicated by stress in a rope linking the space ships. Our example shows the other side of the medal: constant physical distances as determined by radar corresponding to non-zero relative speed.

\subsection{Relative radial velocity and direction of motion}
\label{sec:RelativeRadialDirection}

So far, we have only considered relative speed. What about the direction of relative motion, specifically the sign of the relative radial velocity? The easiest derivation is for the situation where light signals are involved, via the longitudinal Doppler formula (\ref{SRDoppler}): Imagine two successive light wave crests leaving the observer who is at rest at $X=X_1$ at time $T_1$ and $T_1+\Dd T_1$, respectively, and arriving at the observer who is at rest at $X=X_2$ at times $T_2$ and $T_2+\Dd T_2$, respectively. From (\ref{LightPropagationXT}), we have 
\be\Dd T_1=\Dd T_2.
\label{IntervalsEqual}
\ee
That is merely a coordinate statement, though. Physically, each observer measure time intervals on their co-moving clock. Let $\Dd\tau_1$ be the proper time interval observer 1 measures for the interval between the successive departures of the two wave crests, and $\Dd\tau_2$ the proper time measured by observer 2 for the time difference between their arrivals. From (\ref{ProperTimeRelation}), we know
\be
\Dd\tau_{1,2} = \exp(aX_{1,2}/c^2)\cdot\Dd T_{1,2}, 
\ee
so that together with (\ref{IntervalsEqual}), we have
\be
\Dd\tau_1 = \exp(a[X_1-X_2]/c^2)\cdot\Dd\tau_2.
\label{ProperTimeRelation12}
\ee
Given that the wavelength is related to the period as $\lambda=c\cdot \Dd\tau$, (\ref{ProperTimeRelation12}) corresponds to the wavelength shift
\be
1+z = \exp(a[X_2-X_1]/c^2).
\ee
Solving (\ref{SRDoppler}) for $v_R$, we obtain the relative radial velocity
\be
v_R=c\cdot\tanh(a[X_2-X_1]/c^2)
\label{RelativeRadial12}
\ee
for our particular situation, where the events in question are marked by the emission and reception of light. This is consistent with (\ref{LassRelativeSpeed}), but now also includes the overall sign information. 

From this expression for $v_R$, we can derive the second counter-intuitive property of relative motion in relativity. When an observer at constant location $X=X_1$ sends light to an observer at fixed location $X=X_2$, after all, the associated $v_R$ as specified by (\ref{RelativeRadial12}) will have the same magnitude, but the opposite sign than for signals travelling from $X_2$ to $X_1$. The direct consequence in terms of wavelength shifts: Observers will see signals from a source at smaller constant $X$ than their own position as redshifted, and from a source at larger constant $X$ as blueshifted.

Why is that not a logical contradiction? Because in order to determine the relative radial velocity, we always need to specify which two events we are comparing. Of course the result will differ depending on our choice of events. In our example, each observer restricts their analysis to specific event pairs: Each observer only considers events that involve light emitted by the other observer, and received by themselves. These two sets of event pairs are disjunct, so there is no logical contradiction involved if all event pairs in the first set yield a result of ``motion towards,'' while all event pairs in the second set yield a result of ``motion away from.'' It is in this sense that the two observers at rest at two different constant $X$ values are both moving away from each other, $v_R>0$, but also moving towards each other, $v_R<0$. The two seemingly contradictory statements refer to two different sets of event pairs.

\section{Lessons for general relativity}
\label{Sec:Schwarzschild}

The non-intuitive properties of relative velocities described in the previous section can help with understanding certain aspects of general relativity, in particular very basic properties of static gravitational fields.

\subsection{Generalising relative motion}

The definitions of relative speed and relative radial velocity in section \ref{Sec:SRRelativeVelocity} rely heavily on the background structure that is present in special, but not in general relativity: the existence of a family of inertial systems. A natural starting point for generalisation is the formula (\ref{RelativeSpeed}); given two four-velocities $\fvec{w},\fvec{u}$ defined at the same spacetime event, it readily generalises to 
\be
\gamma(v_{\mathrm{rel}}) = -g(\fvec{w},\fvec{u})/c^2,
\label{eq:RelativeSpeedGR}
\ee
where we have replaced the Minkowski metric $\eta$ by the general metric $g$. The challenge is that, in general relativity, this formula can {\em only} be applied to vectors $\fvec{w},\fvec{u}$ that are defined at the same event, mathematically speaking: that are in the same tangent space of our spacetime manifold. But most of the interesting cases involve objects and/or observers whose world lines differ. Here, additional structure linking the different tangent spaces is required:\cite{Synge1960,Narlikar1994,BunnHogg2009} {\em parallel transport} on a mainfold, a procedure which provides us with a prescription for transplanting a four-vector from one tangent space to another, allowing for a direct comparison as via (\ref{eq:RelativeSpeedGR}). In general, the results of parallel transport will depend on the spacetime path along which the vector is transported. That is a direct consequence of non-zero curvature of the spacetime in question.

The notion of parallel transport leads to a natural generalisation of the special-relativistic definition: the {\em relative speed of two objects evaluated along a path that intersects each object's world line} is obtained by taking the first object's four-velocity at one end of the path, parallel-transporting it to the other end of the path, and applying formula (\ref{eq:RelativeSpeedGR}) locally. Scalars such as the scalar product in (\ref{eq:RelativeSpeedGR}) are not changed by parallel transport, so we can perform the comparison at either end of the path. Radial relative velocity can be defined, as well: As evaluated by a locally flat (free-fall) system associated with the receiving object, purely radial motion is when the spatial components of the parallel-transported four-velocity are parallel or anti-parallel to the direction defined by the transport path. For light-like transport paths, there is again the direct link with the wavelength shift of the associated light, and since the comparison takes place in locally flat spacetime, $v_R$ can be recovered directly from the Doppler formula  (\ref{SRDoppler}).\cite{Narlikar1994}

Our departure from defining relative speed as a property of two objects, with no additional information needed, has thus taken us one step further: In classical mechanics, we saw how general motion required us to specify two times, one per object, for the definition of relative speed (or, for purely radial motion, for radial velocity) to work. In special relativity, we need to specify two events, one for each object. In general relativity, in addition we need a path linking the two events. 

At least for events that are sufficiently close to each other, we can eliminate the added arbitrariness
of explicitly choosing a path. As long as the reference events on both objects' world-lines are within a sufficiently small region of space-time, there is a unique straightest-possible line, a geodesic, linking the two events.\cite{ONeill1983} The maximal size of the small region can be estimated from its matter content.\cite{Kannar1992} At least within such a region, we can define relative speed, or radial velocity for purely radial motion, by specifying the two world-lines, and the two events, and then using the unique geodesic joining the two events for parallel transport. Beyond that region, there is ambiguity --- we can end up with more than one relative speed for each pair of objects, and events .

\subsection{Static gravitational fields, gravitational redshift}

Analogs of the two counter-intuitive properties presented in sections \ref{Sec:RelSpeedPhysDistance} and \ref{sec:RelativeRadialDirection} play a role in a particularly interesting situation, namely that of static gravitational fields. The text book example is the Schwarzschild spacetime as the model for both the simplest kind of black hole and, more generally, for the gravitational influence outside any spherically-symmetric mass. The conceptual connection itself should not come as a surprise, given that in the context of the equivalence principle, acceleration can be seen as simulating the simplest possible gravitational field,\cite{Desloge1989,Jones2008,Munoz2010} and that a different set of ``accelerated coordinates,'' namely Rindler coordinates, have long been used as pedagogical tools for exploring black holes, and in particular black hole horizons.\cite{Frolov1998,Faraoni2015,Grumiller2022}

For concreteness, consider the exterior Schwarzschild metric, although the reasoning carries over to general static spacetimes. The original Schwarzschild coordinates are already adapted to the staticity: none of the metric coefficients depends explicitly on time. Consider two static observers, each hovering at a constant $r$ value, one directly below the other. Given the time-independence of the metric, each light signal propagating, say, from the lower to the upper observer takes the same amount of coordinate time to do so. In consequence, two signals sent a coordinate time interval $\Delta t$ apart will also arrive $\Delta t$ apart. In this description, wavelength shifts stem from the different relations between proper time and coordinate time at different $r$ values. The result is a (gravitational) redshift for light travelling outward from the lower to the higher observer, and blueshift for light travelling the other way around.

Using parallel transport to compute the relative radial velocities between the two observers in question, these redshifts can be interpreted as Doppler shifts. A simplified account of this can be found in an article by Narlikar,\cite{Narlikar1994} which in turn is based on the text book of Synge.\cite{Synge1960} Narlikar gives the relative radial motion of a stationary observer at $r_1$ relative to a stationary observer at $r_2$ in a Schwarzschild spacetime with mass parameter $M$, determined by sending light signals from first to the second observer, as
\be
v_R = c\cdot \frac{r_2-r_1}{2r_1r_2/{\cal R}-(r_2+r_1),}
\label{eq:SchwarzschildvR}
\ee
where ${\cal R}\equiv 2GM/c^2$ is the Schwarzschild radius. This and (\ref{SRDoppler}) yield the usual expression for wavelengths shifts for light signals between stationary observers.

For Narlikar, the main focus is on how parallel transport allows for a unified description of Doppler shifts and gravitational redshifts. But there is another aspect: the two apparent contradictions, rooted in classical intuition, that could be used to argue against an interpretation of this situation in terms of relative motion. 

The first apparent contradiction is that the situation involves a static spacetime, and static observers: Both as judged by their constant radial coordinate values and by radar measurements, the two observers are at rest relative to each other, their metric distance as measured along $r$ unchanging. So how can those observers possibly be said to be in relative motion, with non-zero $v_{\mathrm{rel}}$ and $v_R$, as per (\ref{eq:SchwarzschildvR})? Section \ref{Sec:RelSpeedPhysDistance} shows that, even in special relativity, this is no contradiction when accelerated observers are involved, so there is no reason to assume there is a contradiction in general relativity.

That we have a blue-shift for light travelling in one direction, but a red-shift for light travelling in the opposite direction could be taken as a second argument against the consistency of the Doppler interpretation: If those observers {\em are} in relative motion,  surely they must either be moving away from or else towards each other --- so doesn't the asymmetry practically preclude an interpretation as relative motion? Section \ref{sec:RelativeRadialDirection} shows that, on the contrary, this too is a generic property of relative radial velocity for accelerated observers. The key to avoiding a logical contradiction is the same as in our simple model: relative radial velocity is not a property merely of two objects, but also depends on the two events used for the comparison. Blue-shift and red-shift are associated with two different classes of events, one characterised by ingoing, the other by outgoing light-like geodesics.

In these two ways, understanding the properties of relative motion involving accelerations in special relativity can help us understand what is happening in a static gravitational field: by showing certain objections based on our classical intuitions are not applicable in relativity, even in special relativity. 

\subsection{Equivalence principle}

These same properties also provide a new perspective on the equivalence principle. That principle is commonly couched in terms of alternatives: We can interpret a situation as taking place {\em either} with two observers at rest in a gravitational field, {\em or} take the perspective of both observers accelerating, but without the presence of a gravitational field, with the redshift of light travelling from the ``lower'' to the ``higher'' observer interpreted as a gravitational redshift or as a Doppler shift, respectively.

Relativistic relative motion, as applied to a static gravitational field in general relativity, adds a twist: Once the path for parallel transport is fixed, relative radial velocity is an invariant, independent of our chosen perspective. The overall acceleration and the presence of a homogeneous gravitational field might depend on the perspective, but the relative velocity does not. In both cases, there is the same non-zero relative radial velocity, producing an associated Doppler shift. The dichotomy is not between a situation with motion and one with a gravitational field --- relative motion is involved in both cases, and so is the Doppler effect. 

Where classical intuition favours an ``either''--``or,'' either overall accelerated motion or a state of relative rest in a gravitational field, the examples from special relativity show that for accelerated observers, there is on the contrary no way to exclude motion, more specifically relative motion, from the interpretation, even for accelerated observers with a constant (radar) distance.

\subsection{Horizons}

Last but not least, the relative-motion description provides an additional building block for understanding horizons: boundaries separating spacetime into regions whose light can reach us (at least if we wait arbitrarily long) and regions whose light will never reach us. For an external, stationary observer at $r_2$, by (\ref{eq:SchwarzschildvR}), the Schwarzschild horizon at $r_1={\cal R}$ is always at a relative speed $v_R=c$, implying an infinite redshift. Our example situation from section \ref{sec:RelativeRadialDirection} provides a simple analogue for a horizon: For exchanges of light signals between members of our preferred family of observers, see Fig.~\ref{Fig:LassSpacetime}, there is a horizon for light signals sent from $X_1\to -\infty$ to $X_2>X_1$, which by (\ref{RelativeRadial12}) corresponds to $v_R=c$. 

The same intuitive understanding can be applied to cosmological horizons. When the focus in understanding an expanding universe is on the usual recession velocity $v_{\mathrm{rec}}$ taken from the Hubble-Lema\^{\i}tre law, the simple notion that ``behind the cosmological horizons, objects are moving away so fast their light can never reach us, $v_{\mathrm{rec}}>c$'' is incorrect and misleading.\cite{DavisLineweaver2004,Neat2019} Once it is realised that the speed one should consider is the relative radial velocity $v_R$ of Hubble-flow galaxies, several things fall into place. The first is a consistent interpretation of the cosmological redshift as a Doppler shift.\cite{Synge1960,Narlikar1994,BunnHogg2009,Poessel2020b} This $v_R$ is {\em not} the same as the usual $v_{\mathrm{rec}}$. A key difference is that $v_R$ is always sub-luminal, even for the most distant Hubble-flow galaxies we can observe. But the existence of a cosmological horizon, as a boundary between those regions of spacetime in an expanding universe whose light signals can reach us some time in the future and those regions whose light signals can never reach us, is indeed marked by the limit $v_R= c$.\cite{Poessel2020a} Defining relative motion in a suitable way turns a misleading intuition about motion and the light-speed barrier into a helpful one.

\section*{Acknowledgements}
I would like to thank \makeAnon{[anonymized colleague]}{Thomas M\"uller} for his helpful comments on an earlier version of this text. I have no conflicts to disclose.

\end{document}